\documentclass{phbauth}        
\usepackage{graphicx}

\begin{document}

\begin{frontmatter}

\title{Interplay between Structure and Electronic Properties \\
 in Organic Conductors}

\author{Hidetoshi Fukuyama\thanksref{thank1}},
\author{Hitoshi Seo},
\author{Hiori Kino}

\address{Institute for Solid State Physics, University of Tokyo, 
Minato-ku, Tokyo 106-8666, Japan}

\thanks[thank1]{Corresponding author. E-mail: fukuyama@issp.u-tokyo.ac.jp} 

\begin{abstract}
It is known that the ground states of organic conductors have a diversity
reflecting the spatial arrangement of the constituent molecules within the 
unit cell. A systematic theoretical search for the unifying view 
behind such possible ground states has been made based on the 
Hartree-Fock mean field approximation
not only to the on-site but also to intersite Coulomb interactions 
with special emphasis on 
the families of typical organic conductors 
(TMTCF)$_2X$ and (ET)$_2X$. 
\end{abstract}


\end{frontmatter}

\section{Introduction}
The insulating states of solids are realized either
by the absence or the localization of carriers; the
former is the band insulator and the latter cases are
due to either disorder (Anderson localization) or the
mutual Coulomb interactions. 
In the particular case of the non-degenerate half-filled
band the on-site Coulomb interaction leads to the Mott
insulating state, while in the case of the quarter-filled
bands the mutual Coulomb interaction of not only on-site
but also between different sites play important
roles leading to the state where electrons are localized 
every other sites as in the Wigner crystal.
This is a kind of the charge ordering (CO) phenomena. 
In both cases of Mott
insulators and Wigner crystals the electron spins are active
and then the transport and the magnetic properties are closely
interrelated.

In recent years there are much of the research activity
in organic conductors which quite often exhibit metal-insulator
transition as a function of temperature, pressure or magnetic field. 
The experimental studies have elucidated the existence of a diversity
in interesting electronic properties even if the constituent 
molecules and their composition ratio are the same.
Typical examples are (TMTCF)$_2X$ (TMTCF = TMTTF, TMTSF) 
and (ET)$_2X$ (ET = BEDT-TTF).
In these examples, $X$ represent anions attracting almost one 
electron, i.e. $X^-$, and then the conduction processes are 
due to the $\pi$-bands of the donors, TMTCF$^{1/2+}$ or ET$^{1/2+}$
which is quarter-filled if all these molecules are equivalent.
However, in some cases such as quasi-one-dimensional (TMTCF)$_2X$ 
with two donor molecules in a unit cell along the chain direction, 
there exists a small degree of the dimerization, 
which results in neither idealistic quarter-filled nor half-filled bands.
On the other hand in cases of (ET)$_2X$ there often are four ET molecules
in a unit cell with very anisotropic transfer integrals between
molecules, which fact is the cause of the diversity in electronic
properties of these systems, e.g. ground states with antiferromagnetism (AF),
superconductivity (SC), Peierls and spin-Peierls (SP) distortions,  
spin gap (SG) and so on.

In this paper we will briefly review our recent theoretical studies
based on the Hartree-Fock (HF) approximation, 
searching for the simple unifying view behind these apparent complexity.
The effects of quantum fluctuations, which are not taken into account 
in this approximation, are then assessed based on the results thus
obtained.

\section{HF approximation for complex unit cell}

When the electronic state at each site is properly represented 
by one molecular orbital, 
the Hamiltonian will be expressed as follows,
\begin{eqnarray}
H=\sum_{<i,j>}&\sum_{\sigma}&\left(-t_{i,j}a^{\dagger}_{i\sigma}
    a_{j\sigma}+h.c.\right) \nonumber\\
   &+&\sum_{i} U n_{i\uparrow}n_{i\downarrow}
    +\sum_{<i,j>}V_{i,j} n_in_j,
\label{eqn:Hamil}
\end{eqnarray}
Here t$_{i,j}$ is the transfer integrals between molecules 
$i$ and $j$, 
which can be calculated within the extended H\"{u}ckel approximation, 
and $U$ and $V_{i,j}$ are 
on-site and intersite Coulomb interactions, respectively, 
and $n_{is}$ = $a^{\dagger}_{is} a_{is}$ and 
$n_i = n_{i\uparrow} + n_{i\downarrow}$. 
In the actual treatment of the Coulomb interactions 
the $U$ and $V_{i,j}$ terms are 
approximated within the HF and Hartree approximation, 
respectively\cite{Kinopaper}
(we use the term HF below for simplicity).  
%
%
This HF approximation is a durable and systematic procedure for the search of 
possible ground states. 
It should be kept in 
mind, however, that the quantitative aspects of results of this approximation 
should not be taken literally but that the results will be a basis for the further 
detailed theoretical studies.

\section{HF calculations}

\subsection{(TMTCF)$_2X$}\label{TM}
This family has one-dimensional structures 
with the alternation in the transfer 
integrals along the chain direction. 
This dimerization result in a half-filled band, 
instead of the quarter-filling which is expected if all TMTCFs are equivalent.
However it is to be noted that this 
dimerization is very weak compared with the mutual Coulomb interactions as 
will be easily inferred by the smallness of the dimerization gap 
in the band structure, 
and then half-fillingness of this system should be taken with care. 

The experimental phase diagram of this 
family on the plane of pressure and temperature, 
is proposed by J{\'e}rome\cite{Jerome}. 
As the pressure (external as well as chemical) increases, 
the ground state of this system varies as 
SP, commensurate AF (C-AF), incommensurate spin-density-wave (IC-SDW), 
SC and paramagnetic metallic (PM) states. 
An important feature in this phase diagram is the existence of 
temperature T$_{\rho}$, 
where the resistivity shows minimum, 
in the systems whose ground states are SP and C-AF. 
In this low pressure regime, 
the localization of the charge sets in at T$_{\rho}$ as 
the temperature is decreased, independently of the magnetic ordering. 
HF calculations\cite{SeoTM} for the ground states in these cases 
indicate that 
the spatial pattern of the ordered spins in the C-AF phase 
is rather close to that of CO state (as in Wigner crystals) 
where the spins with S=1/2 in every other site,  
rather than a Mott insulator spin S=1/2 on each dimers 
which is the case for strong dimerization. 
The antiferromagnetic pattern in this case, which is schematically 
described as ($\uparrow$ 0 $\downarrow$ 0) 
has actually been observed in some TMTTF compounds\cite{NakamuraTM}. 
The dielectric permitivity measurement\cite{Nad} on (TMTTF)$_2$Br also 
suggest the CO state. 
Moreover, the tencency to CO has been observed directly by NMR 
in different but similar systems 
(DI-DCNQI)$_2$Ag\cite{Hiraki}. 
These results indicate that the origin of the insulating state 
in the region of low pressure
will be a CO state driven by the intersite Coulomb interaction, 
$V$, rather than the simple Mott insulators. 


\subsection{$\kappa$-(ET)$_2X$}
The family of (ET)$_2X$ has many polytypes with a variety of ground state, 
of which the crystal structures are 
basically the same but spatial arrangements of ET molecules in a unit cell 
are different, which result in drastically different ground states as 
described below. 

\begin{figure}[btp]
\begin{center}\leavevmode
\includegraphics[width=0.85\linewidth]{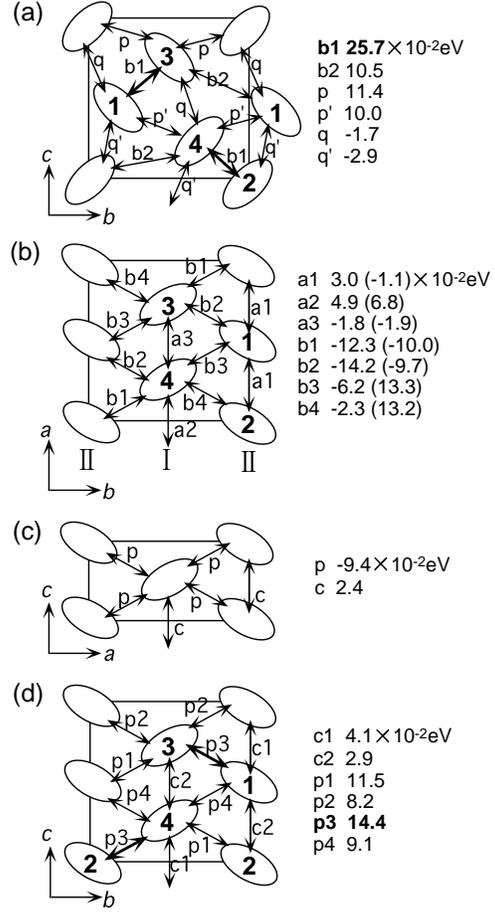}
\caption{Schematic representations of the structures in the donor plane 
for (ET)$_2X$ compounds;  
$\kappa$-type (a), $\theta$-type (b), $\theta_d$-type (c)
(see the text for the notation of $\theta_d$)
and $\alpha$-type (d).
ET molecules are represented as elipses, 
and the intermolecular transfer integrals 
are also shown. (see ref. \cite{Kino,Seotheta})
}
\label{structure}\end{center}\end{figure}

$\kappa$-(ET)$_2X$, which has 4 ETs in a unit cell, 
has the structure shown in Fig. \ref{structure} (a) where 
the transfer integrals between molecules are also indicated. 
This family exhibits a phase diagram 
with an interesting feature of having boundaries between 
AFI and SC\cite{Kanoda}, 
whose symmetry can be different from that of BCS\cite{Kanoda}. 
As seen in Fig. \ref{structure} (a) 
the transfer integral $t_{b1}$ is relatively larger than any others 
implying the formations of the dimers between two molecules connected 
by this $t_{b1}$. Actually, the band structures in the 
paramagnetic state is such that there exists a clear band-gap between 
the higher two and lower two bands,  
which are due to the anti-bonding and bonding orbitals within the dimers, 
respectively. 
The HF calculations reveal that in the presence of large $t_{b1}$,
which is the crucial parameter characterizing 
the degree of dimerization, and of large $U$ 
an AF insulating (AFI) state shown in Fig. \ref{kappa+alpha} (a) 
emerges\cite{Kino}. 
This AFI state is same as the AF 
of spin S=1/2 on each dimer\cite{McKenzie}. 
Once the effective value of $U$, i.e. $U/W$, 
is reduced by pressure, HF calculations predicts 
the transition from AFI to PM, 
which can be SC as in the above case of (TMTCF)$_2X$.

\subsection{$\alpha$-(ET)$_2$I$_3$}
\begin{figure}[btp]
\begin{center}\leavevmode
\includegraphics[width=0.55\linewidth]{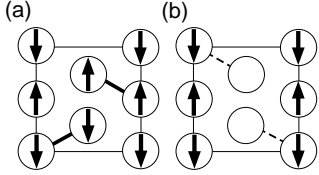}
\caption{A schematic reprensentation of the insulating states 
of $\kappa$-(ET)$_2X$ (a) and $\alpha$-(ET)$_2$I$_3$ (b), 
deduced from HF calculations\cite{Kino}. 
The thick bonds in (a) and the dotted bonds in (b) 
correspond to the transfer integrals 
$t_{b1}$ and $t_{b4}$, repectively. 
}\label{kappa+alpha}\end{center}\end{figure}

The temperature dependence of resistivity of this system is known to
have interesting features, 
i.e. at ambient pressure the resistivity sharply rises 
at around 130 K as $T$ is lowered but with apparently finite limiting value 
towards $T=0$, while at $p=12$ kbar the temperature 
dependence of the resistivity is metallic towards $T=0$\cite{alphares}. 
In the almost insulating state at low temperature at ambient pressure, 
the spin susceptibility is suppressed isotropically 
indicating a non-magnetic ground state with a SG\cite{alphamag}. 
This polytype also has 4 ETs in a unit cell, 
whose spatial arrangement is shown in Fig. \ref{structure} (b).  
For this structure, all of the four bands are weakly separated and the Fermi 
level is located roughly between top-most and 
the next-top bands. 
The HF calculations with only $U$ predict 
that in the case of small values of the transfer integral $t_{b4}$, 
which turns out to be the key factor to determine 
the degree of band overlap between these two bands, 
an AFI state different from that in the $\kappa$-phase, 
as shown in Fig. \ref{kappa+alpha} (b) is stabilized\cite{Kino}. 
This AFI state has 
one-dimensional arrays of spin, whose magnitude is close to S=1/2, 
caused by CO, which we call here `stripes', 
between columns along the $a$-axis. 
However, similar calculations including intersite Coulomb interactions 
predict the stripes along the $b$-axis\cite{Seotheta}. 
This fact indicates that the spatial pattern of CO in this system 
may be very sensitive to the physical parameters.  
Hence the determination realized of this CO is an experimental challenge.

\subsection{$\alpha$-(ET)$_2M$Hg(SCN)$_4$}
\begin{figure}[btp]
\begin{center}\leavevmode
\includegraphics[width=0.7\linewidth]{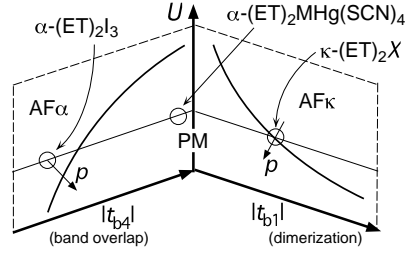}
\caption{A schematic phase diagram of the (ET)$_2X$ 
compounds\cite{Kinopaper}. AF$_{\kappa}$ and AF$_{\alpha}$ are the 
AF states shown in Fig. \ref{kappa+alpha} (a) and (b), respectively. 
Arrows with $p$ denotes the effects of applied pressure. 
}
\label{Kinophd}\end{center}\end{figure}

The spatial arrangement of ETs in this system 
are similar to $\alpha$-(ET)$_2$I$_3$ though the values of transfer integrals 
are different as shown in the parenthesis in Fig. \ref{structure} (b), 
where the value of $t_{b4}$ is quite larger than in the case of 
$\alpha$-(ET)$_2$I$_3$. 
This system is intermediate between $\alpha$-(ET)$_2$I$_3$ and 
$\kappa$-(ET)$_2X$; if the degree of dimerization in $\kappa$-(ET)$_2X$ and 
the degree of the band-overlap in $\alpha$-(ET)$_2$I$_3$ are varied, 
this system is realized\cite{Kinopaper,Kino}. 
This affords a unified view on these three different polytypes,  
as seen in the HF phase diagram shown in Fig. \ref{Kinophd}.

\subsection{$\theta$-(ET)$_2X$}\label{theta}
$\theta$-(ET)$_2MM'$(SCN)$_4$ ($M$=Rb,Cs,Tl, $M'$=Zn,Co) 
are well studied experimentally in recent years, 
and the possibility of CO as the origin of its insulating behavior 
has been proposed\cite{thetaNMR}. 
Depending on the anion [$MM'$(SCN)$_4$]$^-$ or 
even on the cooling rate, 
there are two different arrangements of ET molecules 
in their two-dimensional layers, 
as shown in Fig. \ref{structure} (c) and (d)\cite{Mori}, 
which we call here $\theta$-type and $\theta_d$-type, respectively.  
The $\theta$-type salts has a simply quarter-filled band, 
whereas in $\theta_d$-type salts the bands are split into two 
which are slightly seperated 
due to the weak dimerization: $t_{p3}$ in Fig. \ref{structure} (d) 
is larger than the others. 
HF calculations on each structure suggest various stripe-type CO states 
are stabilized\cite{Seotheta}, where the actual charge pattern 
is sensitive to the values of the parameters of systems.  
We can infer from optical conductivity\cite{Tajima} data
that the stripes are along $c$-axis for $\theta$-type and 
in the $b$-direction along the bonds with $t_{p4}$ 
for $\theta_d$-type structures, respectively, 
which are among the candidates in our results of HF calculations. 


\section{Magnetic properties ---Effects of quantum fluctuation---}
As seen in the previous section, the HF calculations 
indicate the arrangements of spins when the charges are localized. 
However, the quantum fluctuations are neglected 
in the HF calculations, 
and hence these results can be different for some cases 
from the observed magnetic properties. 
Such effects of quantum fluctuations will be properly 
incorporated by mapping the situation to the S=1/2 Heisenberg Hamiltonians 
where these spins are interacting by 
the superexchange interactions deduced from the stuctures. 

\subsection{(TMTCF)$_2X$}
Since the localized state along the chain direction 
in the region of low pressure  
are the CO state as noted in \S \ref{TM}, 
the spin degree of freedom can be represented by 
the quasi-one-dimensional AF Heisenberg model with 
both the coupling to the lattice, $\lambda$, 
and interchain superexchange interaction J$_{\perp}$. 
If the former is dominant, the SP state is realized, but 
as the latter gets important, e.g. by the external pressure, 
it had been theoretically predicted that there will be 
a first order phase transition from the 
SP state to the AF state\cite{Inagaki}, 
which is consistent with the observed properties. 

\subsection{$\kappa$-(ET)$_2X$}
According to the HF calculations, the AFI state of this system 
is as shown in Fig. \ref{kappa+alpha} (a), 
which indicates the existence of (almost) localized spins on 
the triangular lattice with slightly anisotropic 
couplings\cite{Kinopaper,McKenzie}. 
The NMR measurements\cite{Kanoda} show the existence of 
the AF long range order, 
with AF pattern identical to our results, 
which is consistent with the absence of SG in 
calculations on the case of 
the S=1/2 isotropic triangular lattice model. 

\subsection{$\alpha$-(ET)$_2I_3$}
The CO state with the resultant spins along the $a$-axis 
as shown in Fig. \ref{kappa+alpha} (b) 
as well as those along the $b$-axis proposed in ref. \cite{Seotheta} 
will both result in 
a one-dimensional spin system with alternating couplings. 
The magnetic susceptibiliy data seems to suggest the latter case, 
since the magnitude of the superexchange couplings may be 
too small for the former case if we assume that they can be described by 
the simple superexchange relation i.e. $J=4t^2/U$. 

\subsection{$\theta$-(ET)$_2X$}
Similar discussions to the case of $\alpha$-(ET)$_2$I$_3$ 
can be made in this polytype since we found that a stripe type CO 
is also realized in this family. 
In the $\theta$-type compounds, the magnetic susceptibility 
shows a Curie-like behavior without any magnetic ordering 
down to lowest temperatures. 
In contrast, $\theta_d$-phase exhibits a magnetic susceptibility 
with a Bonner-Fisher behavior in the intermediate temperature range 
and below around 10 K, 
a spin gap behavior is observed. 
The stripe type CO with patterns denoted in \S \ref{theta}
are consistent with these measurements, 
since the values of $J=4t^2/U$ with $t=t_{c}$ and $t_{p4}$ 
along the stripes are of the order of few K and 100 K, respectively. 
Then the SG behavior observed in the $\theta_d$-phase 
may be due to the SP transition. 

\subsection{$\lambda$-(BETS)$_2X$}
This family with $X$=GaX$_z$Y$_{4-z}$ has 4 BETS 
(a sulfer analog of ET) molecules in a unit cell as 
in the case of (ET)$_2X$. Experiments have disclosed that the phase diagram 
of this system shares common features with $\kappa$-(ET)$_2X$ 
but with remarkable difference, 
i.e. the insulating state next to the SC state is non-magnetic 
here in contrast to the AFI state in $\kappa$-(ET)$_2X$\cite{Kobayashi}. 
This can be understood theoretically as follows. 
There exists a strong dimerization as in $\kappa$-(ET)$_2X$, 
which will result in a Mott insulating state where S=1/2 are 
localized on each dimer, 
which is inferred from the HF calculations\cite{SeoBETS}.  
The spatial arrangement of BETS molecules
result in the alternation in the resultant superexchange interactions 
along the stacking direction 
together with nonnegligible interchain couplings. 
The phase diagram of such a spin 
system in two-dimension has been studied\cite{Kato}, 
where the ground states are either the AF or 
the SG state. From the actual parameters deduced from 
the structure, it is argued that 
$\lambda$-BETS system is located very close to the boundary between these two  
states, the latter of which is the candidate here\cite{SeoBETS}. 

\section{Summary and Discussion}
The results of recent theoretical studies searching for systematic understanding 
of the variety of ground states realized in (TMTCF)$_2X$ and (ET)$_2X$ salts are 
introduced. Based on the spin dependent Hartree-Fock approximation for the 
Coulomb interaction together with 
the consideration of the full anisotropy of the transfer integrals, 
the key parameters characterizing each family 
(polytype) are extracted resulting in the coherent understanding of the 
apparently unrelated types of the ground states. 
These viewpoints on the organic quarter-filled systems 
are also applicapable to some transition metal oxides 
with quarter-filled band, 
as has been demonstrated in our recent work on NaV$_2$O$_5$\cite{SeoNaV2O5}.

The calculations so far, 
however, have been limited to cases where only one molecular orbital, 
either LUMO or HOMO, is relevant. There exist other interesting cases 
where the overlap between LUMO and HOMO appears to be important 
as in $M$(dmit)$_2$ ($M$=Pd,Ni) compounds or the existence of the 
mixing between $\pi$-orbital and $d$-orbital 
plays crucial roles as in (DCNQI)$_2$Cu and in $\lambda$-(BETS)$_2X$ 
with $X$=FeX$_z$Y$_{4-z}$. 
The studies on the electronic properties of these systems 
with multiple orbital degrees of freedom 
share a common feature with those in the transition metal oxides with 
orbital degeneracy such as LaMnO$_3$ and SrRuO$_4$ systems, 
and these will be a challenging problem in the near feature. 

\begin{ack}
The authors are greatful for K. Kanoda, R. Kato, T. Takahashi 
and many other experimentalists for valuable interactions. 
HS and HK are supported by the JSPS Research Fellowships for 
Young Scientists. 
\end{ack}


\end{document}